\documentclass[sn-mathphys]{sn-jnl}

\usepackage{mathtools}
\DeclarePairedDelimiter{\abs}{\lvert}{\rvert}

\usepackage{bm}

\jyear{2021}%

\theoremstyle{thmstyleone}%
\theoremstyle{thmstyletwo}%

\theoremstyle{thmstylethree}%

\begin{document}

\title[Structural Study of Adsorbed Helium Films]{Structural Study of Adsorbed Helium Films: New Approach with Synchrotron Radiation X-rays}


\author*[1]{\fnm{Akira} \sur{Yamaguchi}}\email{yamagu@sci.u-hyogo.ac.jp}

\author*[2]{\fnm{Hiroo} \sur{Tajiri}}\email{tajiri@spring8.or.jp}

\author[1]{\fnm{Atsuki} \sur{Kumashita}}
\author[2,3]{\fnm{Jun} \sur{Usami}}
\author[1]{\fnm{Yu} \sur{Yamane}}
\author[1]{\fnm{Akihiko} \sur{Sumiyama}}
\author[4]{\fnm{Masaru} \sur{Suzuki}}
\author[5]{\fnm{Tomoki} \sur{Minoguchi}}
\author[2]{\fnm{Yoshiharu} \sur{Sakurai}}
\author[3]{\fnm{Hiroshi} \sur{Fukuyama}}

\affil*[1]{\orgdiv{Graduate School of Science}, \orgname{University of Hyogo}, \orgaddress{\street{3-2-1 Kouto}, \city{Kamigori, Ako}, \postcode{678-1297}, \state{Hyogo}, \country{Japan}}}

\affil*[2]{\orgname{Japan Synchrotron Radiation Research Institute},  \orgaddress{\street{1-1-1 Kouto}, \city{Sayo}, \postcode{679-5198}, \state{Hyogo}, \country{Japan}}}

\affil[3]{\orgdiv{Cryogenic Research Center}, \orgname{The University of Tokyo}, \orgaddress{\street{2-11-16 Yayoi}, \city{Bunkyo-ku}, \postcode{113-0032}, \state{Tokyo}, \country{Japan}}}

\affil[4]{\orgdiv{Department of Engineering Science}, \orgname{University of Electro-Communications}, \orgaddress{\city{Chofu}, \postcode{182-8585}, \state{Tokyo}, \country{Japan}}}
\affil[5]{\orgdiv{Institute of Physics}, \orgname{The University of Tokyo}, \orgaddress{\city{3-8-1 Komaba}, \postcode{153-8902}, \state{Tokyo}, \country{Japan}}}


\abstract{A few atomic layers of helium adsorbed on graphite have been attracting much attention as one of the ideal quantum systems in two dimension. Although previous reports on neutron diffraction have shown fundamental structural information in these systems, there still remain many open questions. Here, we propose surface crystal truncation rod (CTR) scatterings using synchrotron radiation X-rays as a promising method to reveal surface and interface structures of helium films on graphite at temperatures below 2~K, based on the preliminary experimental results on a monolayer of $^4$He on a thin graphite.  Our estimation on heat generation by X-ray irradiations also suggests that CTR scatterings are applicable to even at system temperatures near 100~mK.}




\maketitle

\section{Introduction}\label{sec1}

Adsorbed helium films on graphite are a unique system for studying two dimensional (2D) quantum phenomena \cite{Fukuyama2008,Saunders2020}.  This system has a rich variety of two isotopes ($^3$He and $^4$He, corresponding to fermions and bosons, respectively), different numbers of atomic layers, and various condensed states such as 2D~gas, fluid, and commensurate and incommensurate solids.  Very recently, exotic 2D quantum phases such as  a quantum liquid crystal phase \cite {Nakamura2016} and an intertwined phase like a supersolid \cite{Nyeki2017} have also been proposed to emerge below 1~K. These features,  which are attributed to strong quantum nature of helium atoms and low-dimensionality of the system, have attracted widespread experimental and theoretical attention. Since the appearance of such phases strongly depends on its areal density and layered structure of the helium atoms, structural information is crucial for deeper understandings of them. 

Historically, neutron diffractions were performed from a structural point of view in 1980s to 1990s \cite{Lauter1980,Carneiro1981,Feile1982,Lauter1987,Lauter1990,Lauter1991}, for example,  Lauter et al. investigated in-plane reflections from helium films on partially oriented exfoliated graphite substrates (Pypex and ZYX grade) at temperatures down to 60 mK  \cite{Lauter1990,Lauter1991}.  A large surface area per volume in these exfoliated substrates is advantageous to increase diffraction intensities from them, on the other hand, its wide mosaic spread of graphite microcrystallites makes it difficult to analyze the complete atomic structures within the layers. Currently, the process of layer promotions has been clarified for $^3$He and $^4$He \cite{Lauter1990,Lauter1991}, and is widely accepted as a fundamental knowledge combined with comprehensive heat capacity data by Greywall, et al. \cite{Greywall1990,Greywall1991,Greywall1993}.

However, there are still many open questions, nevertheless neutron diffraction studies are pioneering and have played an important role in this research area. For example, the existence of a commensurate solid in the second layers of $^3$He and $^4$He is still controversy among experiments \cite{Greywall1990,Greywall1993,Nakamura2016} and quantum Monte Carlo calculations \cite{Corboz2008,Gordillo2020}. Therefore, diffraction studies are strongly awaited to extract structural information on these systems.
Here, we propose a new approach to study atomic structures of helium adlayers using synchrotron radiation X-rays. We introduce crystal truncation rod (CTR) scatterings, which have been rapidly developed in recent years as one of the powerful methods to study surface structures. Based on the results of our preliminary study, we discuss the applicability of this method to very low temperatures below 2~K.

\section{Crystal truncation rod scatterings}\label{sec2}

CTR scatterings are known as one of the surface structure analysis techniques~\cite{Tajiri2020}. Since, in general, a crystal is terminated at the surface, its crystallographic symmetry (crystal periodicity) is broken perpendicular to the surface, resulting in appearance of pseudo two-dimensional diffraction conditions in between the Bragg points in the reciprocal space. As a result, additional X-ray scatterings are observed perpendicular to the surface, which is so called the CTR scatterings, as shown in Fig.~1. The intensity of the CTR scatterings $I_{\mathrm {CTR}}(l)$ is formulated in the following equations:

\begin{equation}
\begin{split}
I_{\mathrm{CTR}}(l)&=\abs[\Bigg]{ \sum_{n=- \infty }^{0} F_{hk}^B(l)e^{2\pi inl}+F_{hk}^S(l)}  ^2\\
&=\abs[\Bigg]{ \frac{F_{hk}^B(l)}{1-e^{-2\pi il}}+F_{hk}^S(l)}^2,
\end{split}
\end{equation}

\begin{equation}
F(\bm K)= \int_{\mathrm{cell}}^{}\rho(\bm r )e^{2\pi i K \cdot \bm r}d\bm r.
\end{equation}
Here $F(\bm K)$ is the crystal structure factor, $\rho(\bm r)$ is the electron density in the unit cell at a position  $\bm r$, and $F_{hk}^B(l)$ and $F_{hk}^S(l)$ are a structure factor of the bulk substrate and that of the surface layers, respectively. Here we omitted a prefactor in Eq. (1). A notable point is  that $I_{\mathrm {CTR}}(l)$ is not proportional to an individual sum of the contributions of bulk substrate and the surface layers, but to a squared absolute value of the sum of them. Since both $F_{hk}^B(l)$ and $F_{hk}^S(l)$ are complex numbers, interference is expected among them. From Eq. (1), it is derived that the bulk and surface contributions are nearly equivalent at points with fractional $l$ in the reciprocal space.  Consequently, the CTR scattering is quite sensitive to the adsorbed layers on the surface.  Furthermore, by combining this technique with intense and highly parallel synchrotron radiation X-rays, it is possible to study the structure of helium films in spite of their small scattering factors.

\begin{figure}[b]
\begin{center}
  \includegraphics[width=0.8\textwidth]{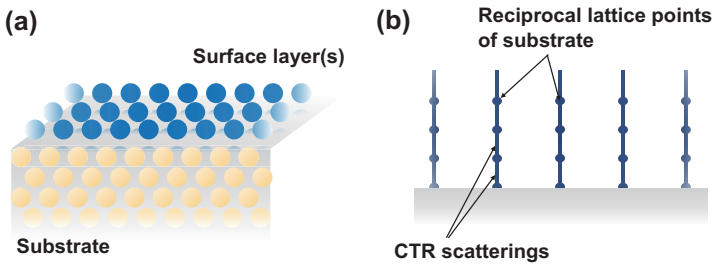}
\caption{ (Color online) Schematic views of a substrate surface with an adlayer in the real space (a), and in the reciprocal space (b). }
\label{fig:1}       
\end{center}
\end{figure}

In this work, we focus on scatterings from the 00~rod (or $hkl = 00l $ in the Miller indices), which contain surface structural information perpendicular to the surface. As mentioned above, the previous neutron studies have focused on in-plane structure only using in-plane reflections, except the one \cite{Carneiro1981}.  Our approach will provide us with new information on heights of helium adlayers.

\section{Preliminary experiment}\label{sec3}

\begin{figure}[b]
\begin{center}
  \includegraphics[width=0.75\textwidth]{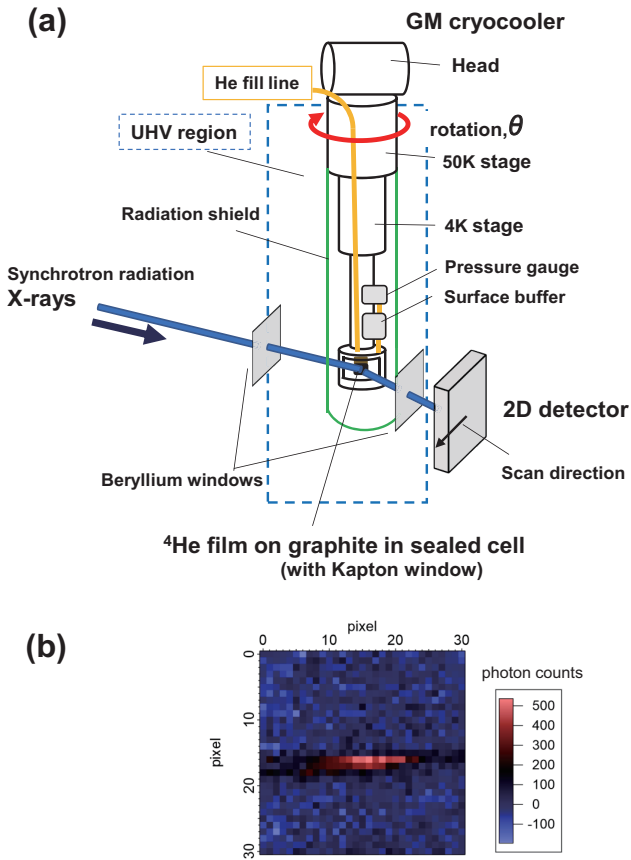}
\caption{(Color online) (a) Schematic view of the cryogenic part of the experimental setup. It is installed in the UHV chamber, represented by the blue dashed line. (b) Color plot of CTR scatterings with 00$L$ ($L$=1.66) for a $^4$He monolayer on graphite sample. The horizontal axis of the image is parallel to the surface normal. The pixel size is 100$\times$100~$\mu$m$^2$. The color bar indicates photon counts at each pixel.  A background scattering from misaligned microcrystallites which lies along the scan direction has already been subtracted. }
\label{fig:2}       
\end{center}
\end{figure}

In order to evaluate feasibility of the X-ray diffraction experiment we suggest, we have preformed a preliminary study at the surface and interface strucutres beamline (BL13XU \cite{Sakata2003,Tajiri2019}) of SPring-8 in Japan. Figure 2(a) shows a schematic of the cryogenic part of the experimental setup. The refrigerator, which is based on a GM cryocooler (Sumitomo Heavy Industries, Ltd., RDK-101E), is designed to match the ultra-high-vacuum (UHV) chamber in the beamline.  The refrigerator rotates 360$^\circ$ around the vertical axis, allowing us to perform, so called,  $\theta$ scan. The bottom part of the refrigerator is equipped with a 1~K pot, which is used to cool the system further down to 1.6~K. A demountable sample cell was attached at the bottom of the 1~K pot, which is helium-leak tight with a large window made of a thin Kapton wall. The window is designed to let incident and reflected X-ray beams pass through it with reduced absorption. Beryllium X-ray windows for incident and reflected X-rays are equipped with the UHV chamber.  Scatterings and absorption at the beryllium windows as well as aluminized-Kapton-film radiation shields attached to the 50~K stage were also sufficiently small. The scattering data were collected with a 2D detector.

We used an highly oriented pyrolitic graphite (HOPG) thin film of high quality as a graphite substrate in this experiment. The HOPG substrate with 50~$\mu$m thickness, which is fabricated on a 1.7~mm thick glass plate, was adhered to a small plate of copper with sliver paste from the graphite side, and the copper plate was thermally anchored to 1~K pot. A notch was machined into the copper plate as  a loophole of X-ray beams. Details of the substrate will be described elsewhere \cite{Tajiri2021}. Since the surface area of the HOPG is small, we use a surface buffer made of Grafoil with a surface area of 5.4~m$^2$ in order to control the areal density and number of layers of $^4$He. A cryogenic pressure gauge was installed to monitor the adsorption pressure during sample preparations.

Figure 2(b) shows a CTR scattering image at 00$L$ ($L=1.66$) for $^4$He monolayer on graphite obtained at 4.5~K. At this temperature, the $^4$He film is in the fluid phase \cite{Greywall1991,Greywall1993}. X-rays of 20~keV in energy with the beam size of 0.3 mm in diameter, which corresponds to photon flux of 2$\times$10$^{11}$ photons/s, were used. The measurement time duration per each image was 10~s. In this condition, our measurement was successfully performed without significant heat generation, where the integrated photon counts of reflections were  above 1000, indicating that the statistical error of each measurement is less than 4\%.

\section{Results and discussion}\label{sec4}

The major concern in cryogenic scattering experiments using synchrotron radiation X-rays will be the heat generation by X-ray irradiation.  From the preliminary study, the photon flux of $2\times 10^{11}$ photons/s was found to be enough to obtain sufficient CTR scattering intensity. Here we evaluate a temperature increase of the $^4$He film with this irradiation intensity, assuming a simplified geometry, as shown in Fig.~3.  In the geometry, the X-ray  beam with 0.3~mm in diameter enters an circular HOPG substrate of 50~$\mu$m thick and 1~mm in diameter, surrounded by a copper plate of 50~$\mu$m thick. We assume the perfect thermal contact between the HOPG film and the copper plate. The glass supporting plate is omitted in the estimation for simplicity and the effect is discussed later. The temperature of the copper plate, $T_{\mathrm{bath}}$ is fixed (constant) as a thermal bath.  From the photon energy (20~keV) and flux intensity ($2\times10^{11}$~photons$/$s), the total power of the irradiation X-ray is calculated to be $640~\mu$W. Note that, since the incident X-rays are in the hard X-ray region, most of them pass through the graphite substrate film, and only 0.4\% of the X-rays are absorbed in a 50~$\mu$m thick graphite film \cite{Xraydatabase}. Furthermore, if the mass energy-transfer coefficient \cite{Kato2014} is taken into account, only about 0.2\%, namely, 1.3~$\mu$W of the energy is actually converted to heat in the graphite substrate. This heat is deposited on the red area in Fig. 3, and diffuses to the surroundings by heat conduction through the HOPG.  First, we estimate the temperature rise at $T_{\mathrm{bath}}=1~$K.  The thermal conductivity of HOPG is highly anisotropic; the in-plane and out-of-plane thermal conductivities are  $\kappa_{\parallel}=1.0$ \cite{Morelli1985} and $\kappa_{\perp}=1\times 10^{-2}$~\cite{Uher1985}~W/(K$\cdot$m) at 1~K, respectively. Because of the high in-plane thermal conductivity, the increase of temperature is only $\Delta T= 4.9$~mK at the graphite surface of the irradiation side, namely, at the position of the helium film. The result indicates that the CTR scattering experiment is feasible at $T=1$~K with $\Delta T/T=0.49$\%. 

Similarly, at $T_{\mathrm{bath}}=100$~mK where $\kappa_{\parallel}=6\times10^{-2}$ \cite{Morelli1985} and $\kappa_{\perp}=2\times 10^{-4}$~\cite{Uher1985}~W/(K$\cdot$m), we estimate the temperature increase is 82~mK. This temperature increase can be reduced by one order of magnitude by deceasing the beam intensity and increasing the measurement time. For example, for  a photon flux of $1\times 10^{10}$ photons/s and measurement time of 200~sec, it will be $\Delta T=4.1$~mK at 100~mK ($\Delta T/T=4.1$\%).

\begin{figure}[t]
\begin{center}
  \includegraphics[width=0.8\textwidth]{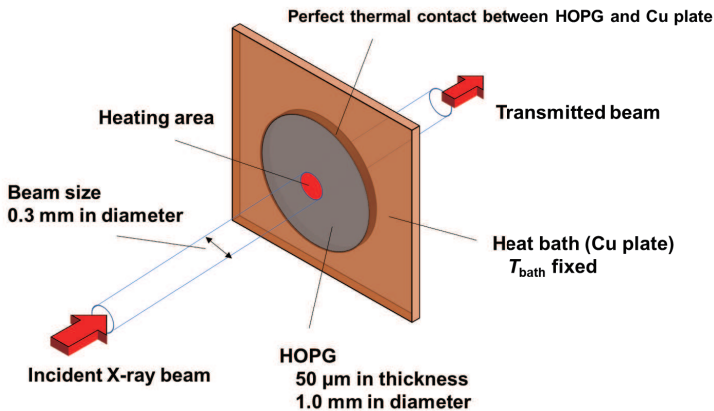}
\caption{(Color online) Thermal model for estimation of temperature rise at at the graphite surface. See the text for details.}
\label{fig:3}       
\end{center}
\end{figure}

In reality, the 1.7~mm thick glass plate, which was attached on the back side of the HOPG film in the preliminary experiment, should have absorbed a large amount of X-rays and generated an enormous heat that is unbearable when the temperatures go down to subkelvin. However, this problem can be managed by shaving the glass plate mechanically down to about 100~$\mu$m thick.  It is noted that this extra heat generation can be eliminated by using self-standing single crystal graphite, instead of the thin HOPG film.  From the above discussion, we conclude that the CTR scattering experiment is quite feasible at least down to $T\sim$100~mK.

Finally, we summarize the advantages of the proposed CTR scatterings. Compared with the conventional neutron diffraction using exfoliated graphite whose microcrystallites have wide spreads of mosaic angle and orientation, CTR scatterings are very effective to extract structural information on surface systems including helium layers on graphite by using a single crystal with very small mosaic spread. In addition, X-ray beam size can be reduced to a few micrometers in diameter at the state-of-the-art synchrotron facilities like SPring-8, which is applicable to phase separation (or coexisting) phenomena with enough spatial resolution, such as the gas-liquid phase separation in 2D $^3$He, one of interesting topics proposed recently \cite{Sato2012}. Furthermore, the height information for each atomic layer in He multilayers has not been clarified so far, and, thus, our approach using CTR scatterings offers novel experimental insight into them.  Comparison between our expected structural data and the quantum Mote Carlo simulations, which have become progressively accurate in recent years \cite{Corboz2008, Gordillo2020,Gordillo2012}, will produce fruitful discussion or novel findings.

\section{Conclusion}\label{sec5}

We proposed a new approach for studying structures of 2D helium films on graphite using CTR scatterings with synchrotron radiation X-rays. In our preliminary study, we succeeded in detecting CTR scattering clearly from monolayer of $^4$He films adsorbed on a thin HOPG surface at 4.5 K. From our estimation on influence of heat, we concluded that CTR scattering measurements are possible even in temperatures down to near 100~mK.

The datasets generated during and/or analysed during the current study are available from the corresponding author on reasonable request.

\backmatter

\bmhead{Acknowledgments}

We thank Dr. Mototada Kobayashi for providing HOPG samples, and Prof. Gaku Motoyama, Prof. Takashi Nishioka and Prof. Ryuji Nomura for helpful discussions of the refrigerator design.
This work was partly supported by JSPS KAKENHI Grant Numbers JP18H01170, JP18H03479, JP20H05621, and a special research grant of University of Hyogo. J.U. was supported by JSPS through MERIT program and JSPS Fellows (JP20J12304).
The surface X-ray scattering measurements were
performed using BL13XU, SPring-8 with the approval of the Japan Synchrotron
Radiation Research Institute (Proposal No.  2020A0599, 2020A2045, 2020A2137, 2021A1142, 2021A2070).




\end{document}